\documentclass[floatfix,showpacs,aps,prb,twocolumn]{revtex4}
\usepackage[utf8]{inputenc}
\usepackage[T1]{fontenc}
\usepackage{bm}
\usepackage{makeidx}
\usepackage{eurosym}
\usepackage{graphicx}
\usepackage{graphics}
\usepackage{amssymb}
\usepackage{amsmath}
\usepackage{color}
\usepackage{float}
\usepackage{hyperref}
\usepackage{fancyhdr}
\usepackage{booktabs}
\usepackage{stackengine}

\DeclareUnicodeCharacter{03B3}{$\gamma$}

\begin{document}

\title{On the origin of Field-Induced Boson Insulating States in a 2D
Superconducting Electron Gas with Strong Spin-Orbit Scatterings }
\author{Tsofar Maniv}
\author{Vladimir Zhuravlev}
\affiliation{Schulich Faculty of Chemistry, Technion-Israel Institute of Technology,
Haifa 32000, Israel}
\email{maniv@technion.ac.il}
\date{\today }

\begin{abstract}
We search for the deep origin of the field-induced
superconductor-to-insulator transitions observed experimentally in
electron-doped SrTiO$_{3}$/LaAlO$_{3}$ interfaces, which were analyzed
theoretically very recently within the framework of superconducting
fluctuations approach (Phys. Rev. B \textbf{104}, 054503 (2021)). Employing
the 2D electron-gas model with strong spin-orbit scatterings, we have found
that in the zero temperature limit, field-induced unbounded growth of the
fluctuation mass, and consequent divergence of Cooper-pair density in
mesoscopic puddles, drives the system to Boson insulating states at high
fields. Application of this model to the gate-voltage tuned 2D electron
system, created in the SrTiO$_{3}$/LaAlO$_{3}$ (111) interface at low
temperatures, shows that, at sufficiently high fields, the DOS conductivity
prevails over the paraconductivity, resulting in strongly enhanced
magnetoresistance in systems with sufficiently small carriers density.
Dynamical quantum tunneling of Cooper pairs breaking into mobile
normal-electrons states, which prevent the divergence at zero temperature,
contain the high-field resistance onset. 
\end{abstract}

\maketitle

In a very recent paper \cite{MZPRB2021} we have shown that Cooper-pair
fluctuations in a 2D electron gas with strong spin-orbit scatterings can
lead at low temperatures to pronounced magnetoresistance (MR) peaks above a
crossover field to superconductivity. The model was applied to the high
mobility electron systems formed in the electron-doped interfaces between
two insulating perovskite oxides---SrTiO$_{3}$ and LaAlO$_{3}$ \cite%
{Ohtomo04}, showing good quantitative agreement with a large body of
experimental sheet-resistance data obtained under varying gate voltage \cite%
{Mograbi19}.

The model employed was based on the opposing effects generated by
fluctuations in the superconducting (SC) order parameter: The nearly
singular enhancement of conductivity (paraconductivity) due to fluctuating
Cooper pairs below the nominal (mean-field) critical magnetic field, on one
hand, and the suppression of conductivity, associated with the loss of
unpaired electrons due to Cooper pairs formation, on the other hand. The
self-consistent treatment of the interaction between fluctuations \cite%
{UllDor90},\cite{UllDor91}, employed in these calculations, avoids the
critical divergence of both the Aslamazov-Larkin (AL) paraconductivity \cite%
{AL68} and the DOS conductivity \cite{LV05}, allowing to extend the theory
to regions well below the nominal critical SC transition. The absence of
long range phase coherence implied by this approach is consistent with the
lack of the ultimate zero-resistance state in the entire data analyzed there.

In the present paper we focus our attention on the most intriguing question
arising from the Cooper-pair fluctuations scenario of the
superconductor--insulator transition (SIT) presented in Ref.\cite{MZPRB2021}%
, that is how Cooper-pairs liquid, whose condensation (in momentum space) is
customarily associated with superconductivity, could metamorphose into an
insulator just by lowering its temperature under sufficiently high magnetic
field ? We have already identified the highly suppressed normal-state DOS
due to Cooper-pairs formation as the dominant origin of the insulator side
of the SIT.

Here we show that field-induced vanishing of the fluctuations stiffness in
the zero temperature limit is at the core of this intriguing phenomenon.
Under these extreme circumstances, the fluctuation mass enhances without
limit, the AL paraconductivity vanishes and the DOS conductivity diverges,
so that at low but finite temperature the DOS conductivity prevails over the
AL conductivity at fields that roughly indicate the presence of the observed
enhanced MR.

It is therefore concluded that the consequent divergence of the Cooper-pairs
density within mesoscopic puddles, as predicted by the thermal fluctuations
approach in the zero temperature limit, should bolster dynamical quantum
tunneling of Cooper pairs breaking into unpaired{\Large \ }mobile electrons
states, and so containing the resistance onset. This feature reflects on the
overall comparison process with the experimental data, which shows selective
sensitivity to the phenomenological parameters determining both the rate of
quantum tunneling and the normal-state conductivity. \bigskip

\section{Conductance fluctuations in the zero temperature limit}

In order to reveal the origin of the puzzling insulating state that emerges
in our approach from SC fluctuations we will consider in this section the
fluctuations contributions to the sheet conductivity in the magnetic fields
region where they are rigorously derivable from the microscopic Gor'kov
Ginzburg-Landau theory, i.e. above the nominal (mean-field) critical field,
determined from the vanishing of the Gaussian critical shift-parameter \cite%
{MZPRB2021}: 
\begin{equation}
\varepsilon _{H}\equiv \ln \left( \frac{T}{T_{c0}}\right) +a_{+}\psi \left( 
\frac{1}{2}+f_{-}\right) +a_{-}\psi \left( \frac{1}{2}+f_{+}\right) -\psi
\left( 1/2\right)  \label{eps_H}
\end{equation}

Here $T_{c0}$ is the mean-field SC transition temperature at zero magnetic
field, $\psi $ is the digamma function, $\ f_{\pm }=\delta H^{2}+\beta \pm 
\sqrt{\beta ^{2}-\mu ^{2}H^{2}}$, $a_{\pm }=\left( 1\pm \beta /\sqrt{\beta
^{2}-\mu ^{2}H^{2}}\right) /2$ are dimensionless functions of the magnetic
field $H$, with the basic parameters: $\ \beta \equiv \varepsilon _{SO}/4\pi
k_{B}T,$ $\mu \equiv \mu _{B}/2\pi k_{B}T,$ $\delta \equiv D\left( de\right)
^{2}/2\pi k_{B}T\hslash $, where $D\equiv \hbar E_{F}/m^{\ast }\varepsilon
_{SO}$ the electron diffusion coefficient, and $\varepsilon _{SO}=\hbar
/\tau _{SO}$ is the spin-orbit energy. There are no restrictions on the
temperature $T$ as we are mainly interested in the low temperatures region
well below $T_{c0}$ down to the limit of $T\rightarrow 0$.

\subsection{DOS conductivity}

As indicated in \cite{MZPRB2021}, the phenomenological approach to the
calculation of the DOS conductivity, based on the simple Drude formula: $%
\delta \sigma _{DOS}=-2n_{s}e^{2}\tau _{SO}/m^{\ast }$, as first introduced
by Larkin and Varlamov \cite{LV05} for the zero-field case in the dirty
limit, can fit nicely the result derived by means of a fully microscopic
(diagrammatic) approach. The key factor is the Cooper-pair fluctuations
number density $n_{s}$:

\begin{equation}
n_{s}=\frac{1}{d}\frac{1}{\left( 2\pi \right) ^{2}}\int \left\langle
\left\vert \psi \left( q\right) \right\vert ^{2}\right\rangle d^{2}q
\label{n_s}
\end{equation}%
which depends on an appropriate selection of its momentum distribution
function: $\left\langle \left\vert \psi \left( q\right) \right\vert
^{2}\right\rangle $. \ The latter was selected by generalizing the
pure-limit zero-field expression \cite{LV05}: $\left\langle \left\vert \psi
\left( q\right) \right\vert ^{2}\right\rangle =\alpha ^{-1}\left[ \ln \left(
T/T_{c0}\right) +\xi ^{2}q^{2}\right] ^{-1},$ with:$\ \alpha =4\pi
^{2}k_{B}T/7\zeta \left( 3\right) E_{F}$, $\zeta \left( 3\right) \simeq
1.202 $, and $\xi =\hbar v_{F}/2\pi k_{B}T$, to the dirty-limit finite-field
expression:

\begin{equation}
\left\langle \left\vert \psi \left( q\right) \right\vert ^{2}\right\rangle
\simeq \frac{7\zeta \left( 3\right) E_{F}}{4\pi ^{2}k_{B}T}\frac{1}{\Phi
\left( x;\varepsilon _{H}\right) }  \label{MomDistr}
\end{equation}%
where:

\begin{eqnarray}
\Phi \left( x;\varepsilon _{H}\right) &=&\varepsilon _{H}+a_{+}\left[ \psi
\left( 1/2+f_{-}+x\right) -\psi \left( 1/2+f_{-}\right) \right]  \notag \\
&&+a_{-}\left[ \psi \left( 1/2+f_{+}+x\right) -\psi \left( 1/2+f_{+}\right) %
\right]  \label{Phi(x)}
\end{eqnarray}%
and $x=\hbar Dq^{2}/4\pi k_{B}T$. \ The resulting expression of the DOS
conductivity contribution is given by: 
\begin{equation}
\delta \sigma _{DOS}d\simeq -3.5\zeta \left( 3\right) \left( \frac{G_{0}}{%
\pi }\right) \int_{0}^{x_{c}}\frac{dx}{\Phi \left( x;\varepsilon _{H}\right) 
}  \label{delsig_DOSd}
\end{equation}%
where $G_{0}=e^{2}/\pi \hbar $ is the conductance quantum, $x_{c}=\hbar
Dq_{c}^{2}/4\pi k_{B}T$, with $q_{c}$ the cutoff wave number, and $3.5\zeta
\left( 3\right) \simeq 4.207$.

Further insight into the zero-temperature limit of $\delta \sigma _{DOS}d$
is gained by exploiting the linear approximation of Eq.(\ref{Phi(x)}), i.e.:
\ $\Phi \left( x;\varepsilon _{H}\right) \simeq \varepsilon _{H}+\eta \left(
H\right) x$, where:

\begin{equation}
\eta \left( H\right) =a_{+}\psi ^{\prime }\left( \frac{1}{2}+f_{-}\right)
+a_{-}\psi ^{\prime }\left( \frac{1}{2}+f_{+}\right)  \label{eta}
\end{equation}%
and performing the integration over $x$ analytically, which yields: 
\begin{equation}
\delta \sigma _{DOS}d\simeq -3.5\zeta \left( 3\right) \left( \frac{G_{0}}{%
\pi }\right) \frac{1}{\eta \left( H\right) }\ln \left( 1+\frac{\eta \left(
H\right) x_{c}}{\varepsilon _{H}}\right)  \label{delsig_DOSdlin}
\end{equation}

In the zero field limit ( $\eta \left( H\rightarrow 0\right) =\psi ^{\prime
}\left( 1/2\right) =\pi ^{2}/2\equiv \eta $), and for sufficiently large
cutoff, i.e. $x_{c}\gg \varepsilon _{H}/\eta \left( H\right) $, we find: \ $%
\delta \sigma ^{DOS}\left( H\rightarrow 0\right) \simeq -3.5\zeta \left(
3\right) \left( \frac{G_{0}}{\pi d\eta }\right) \ln \frac{\eta x_{c}}{%
\varepsilon }$, so that:

\begin{equation*}
\delta \sigma ^{DOS}\simeq -\left( \frac{7\zeta \left( 3\right) }{\pi ^{4}}%
\right) \left( \frac{e^{2}}{d\hbar }\right) \ln \left( \frac{\eta x_{c}}{%
\varepsilon }\right)
\end{equation*}%
in complete agreement \cite{footnote1} with the result of a fully
microscopic (diagrammatic) approach presented in Appendix A (Eq.\ref%
{sigDOS_LV}) for a 2D system, following the method used in Ref.\cite{LV05}
for a layered superconductor. \ 

\subsection{Paraconductivity}

The AL contribution to the sheet conductance derived in Ref.\cite{MZPRB2021}
was obtained from the retarded current-current correlator: 
\begin{eqnarray}
&&Q_{AL}^{R}\left( \omega \right) =k_{B}T\left( \frac{2e}{\hbar }\right)
^{2}\left( \frac{1}{2\pi d}\right) \int\limits_{0}^{x_{c}}xdx  \label{Q_AL^R}
\\
&&\sum\limits_{n=0,\pm 1,\pm 2,....}\frac{\Phi ^{\prime }\left( x+\left\vert
n+y\right\vert ;\varepsilon _{H}\right) }{\Phi \left( x+\left\vert
n+y\right\vert ;\varepsilon _{H}\right) }\frac{\Phi ^{\prime }\left(
x+\left\vert n\right\vert ;\varepsilon _{H}\right) }{\Phi \left(
x+\left\vert n\right\vert ;\varepsilon _{H}\right) }  \notag
\end{eqnarray}%
where $y=i\hbar \omega /2\pi k_{B}T$, and $\omega $ is the frequency of the
response function.

Using Eq.\ref{Q_AL^R} all nonzero Matsubara-frequency terms in the
corresponding AL static conductivity $\sigma _{AL}=\lim_{\omega \rightarrow
0}\left( i/\omega \right) \left[ Q_{AL}^{R}\left( \omega \right)
-Q_{AL}^{R}\left( 0\right) \right] $ are canceled out and the remaining $%
n=0 $ term can be written in the form: 
\begin{equation}
\sigma _{AL}d=\frac{1}{4}\left( \frac{G_{0}}{\pi }\right)
\int\limits_{0}^{x_{c}}\left( \frac{\Phi ^{\prime }\left( x;\varepsilon
_{H}\right) }{\Phi \left( x;\varepsilon _{H}\right) }\right) ^{2}dx
\label{sig_ALd}
\end{equation}

Exploiting the linear approximation of Eq.(\ref{Phi(x)}), i.e.: \ $\Phi
\left( x;\varepsilon _{H}\right) \simeq \varepsilon _{H}+\eta \left(
H\right) x$, and performing the integration over $x$ analytically we find:

\begin{equation}
\sigma _{AL}d\simeq \frac{1}{4}\left( \frac{G_{0}}{\pi }\right) \frac{\eta
\left( H\right) }{\varepsilon _{H}\left( 1+\frac{\varepsilon _{H}}{\eta
\left( H\right) x_{c}}\right) }  \label{sig_ALdlin}
\end{equation}

\subsection{Infinite boson mass at zero temperature}

Combining Eq.(\ref{delsig_DOSdlin}) with Eq.(\ref{sig_ALdlin}), the total
fluctuations contributions to the sheet conductance is written as:

\begin{eqnarray}
&&\sigma ^{fluct}d\simeq \left( \frac{G_{0}}{\pi }\right) \left[ \frac{\eta
\left( H\right) }{4}\frac{1}{\varepsilon _{H}\left( 1+\frac{\varepsilon _{H}%
}{\eta \left( H\right) x_{c}}\right) }\right.  \notag \\
&&\left. -\frac{3.5\zeta \left( 3\right) }{\eta \left( H\right) }\ln \left(
1+\frac{\eta \left( H\right) x_{c}}{\varepsilon _{H}}\right) \right]
\label{sig^fluctdlin}
\end{eqnarray}%
which highlights the complementary roles played by the stiffness parameter $%
\eta \left( H\right) $ in the AL and DOS conductivities. The importance of $%
\eta \left( H\right) $ in controlling the development of an insulating
bosonic state at low temperatures and high magnetic field can be clearly
understood by considering the extreme situation of its zero temperature
limit.

To effectively investigate this limiting situation it will be helpful to
rewrite $\eta \left( H\right) $ as a sum over fermionic Matzubara frequency,
that is: 
\begin{equation}
\eta \left( h\right) =\sum\limits_{n=0}^{\infty }\frac{\left( n+1/2+2\beta +%
\overline{\delta }h^{2}\right) ^{2}-\overline{\mu }^{2}h^{2}}{\left[ \left(
n+1/2+\overline{\delta }h^{2}\right) \left( n+1/2+2\beta +\overline{\delta }%
h^{2}\right) +\overline{\mu }^{2}h^{2}\right] ^{2}}  \label{eta(h)}
\end{equation}
where: $h\equiv H/H_{c\parallel 0}^{\ast },t\equiv T/T_{c}^{\ast },\beta
=\beta _{0}/t,\overline{\mu }=\mu _{0}/t,\overline{\delta }=\delta
_{0}/t,\beta _{0}\equiv \varepsilon _{SO}/4\pi k_{B}T_{c}^{\ast },\delta
_{0}\equiv D\left( deH_{c\parallel 0}^{\ast }\right) ^{2}/2\pi
k_{B}T_{c}^{\ast }\hslash ,$

$\mu _{0}\equiv \mu _{B}H_{c\parallel 0}^{\ast }/2\pi k_{B}T_{c}^{\ast }$,
with $H_{c\parallel 0}^{\ast }$ and $T_{c}^{\ast }$ being characteristic
scales of the critical parallel magnetic field and critical temperature,
respectively.

In the zero temperature ($t\rightarrow 0$) limit, at finite magnetic field, $%
h>0$, the discrete summation in Eq.(\ref{eta(h)}) transforms into
integration, i.e.: $\ \eta \left( h\right) \rightarrow
t\int\limits_{0}^{\infty }d\nu \frac{\left( \nu +2\beta _{0}+\delta
_{0}h^{2}\right) ^{2}-\mu _{0}^{2}h^{2}}{\left[ \left( \nu +\delta
_{0}h^{2}\right) \left( \nu +2\beta _{0}+\delta _{0}h^{2}\right) +\mu
_{0}^{2}h^{2}\right] ^{2}}=t\frac{1}{h^{2}}\frac{\delta _{0}h^{2}+2\beta _{0}%
}{h^{2}\delta _{0}^{2}+2\beta _{0}\delta _{0}+\mu _{0}^{2}}$, so that:

\begin{equation}
\eta \left( h\right) \rightarrow t\left( \frac{\eta _{0}\left( h\right) }{%
h^{2}}\right) \rightarrow 0  \label{eta(h)t0}
\end{equation}%
where: 
\begin{equation}
\eta _{0}\left( h\right) \equiv \frac{\delta _{0}h^{2}+2\beta _{0}}{\left(
\delta _{0}h^{2}+2\beta _{0}\right) \delta _{0}+\mu _{0}^{2}}
\label{eta_0(h)}
\end{equation}

Note that at zero magnetic field: $\eta \left( h=0\right)
=\sum_{n=0}^{\infty }\left( n+1/2\right) ^{-2}=\psi ^{\prime }\left(
1/2\right) =\pi ^{2}/2$, independent of temperature.

Thus, the zero temperature limit of the sheet conductance, Eq.(\ref%
{sig^fluctdlin}), at fields above the nominal critical field $H_{c\parallel
0}^{\ast }$ can be written in the form: 
\begin{align}
& \left( \sigma ^{fluct}\right) _{h>1,t\rightarrow 0}d\rightarrow \left( 
\frac{G_{0}}{\pi }\right) \left[ t\left( \frac{\eta _{0}\left( h\right) }{%
4h^{2}}\right) \frac{1}{\varepsilon _{h}\left( 1+\frac{h^{2}\varepsilon _{h}%
}{\eta _{0}\left( h\right) x_{0}}\right) }\right.  \notag \\
& \left. -\frac{1}{t}\left( \frac{3.5\zeta \left( 3\right) h^{2}}{\eta
_{0}\left( h\right) }\right) \ln \left( 1+\frac{\eta _{0}\left( h\right)
x_{0}}{h^{2}\varepsilon _{h}}\right) \right]  \label{sig^fluctdt0}
\end{align}%
where $x_{0}\equiv \hbar Dq_{c}^{2}/4\pi k_{B}T_{c}^{\ast }$ is the
temperature-independent cutoff parameter{\normalsize . }It should be
stressed at this point that the temperature-independent argument of the
logarithmic factor in Eq.(\ref{sig^fluctdt0}) (see Ref.\cite{footnote2}) is
consistent with the temperature-dependent cutoff parameter {\normalsize $%
x_{c}=x_{0}/t$.}

Thus, we conclude that in the $t\rightarrow 0$\ limit the AL
paraconductivity follows the vanishing stiffness parameter $\eta \left(
h\right) \propto t$, Eq.(\ref{eta(h)t0}), whereas the DOS conductivity
diverges with $1/\eta \left( h\right) \propto 1/t$. The former effect is a
direct consequence of the divergent effective mass of the fluctuations,
whereas the latter is due to the unlimited accumulation of Cooper-pairs
within fluctuation puddles, whose characteristic spatial size:

\begin{equation*}
\xi \left( t\rightarrow 0\right) =\left( \frac{\eta _{0}\left( h\right) }{%
h^{2}\varepsilon _{h}}\frac{\hbar D}{4\pi k_{B}T_{c}^{\ast }}\right) ^{1/2}
\end{equation*}%
remains finite in this extreme limiting situation. The decreasing asymptotic
field dependence ($\eta \left( h\right) \propto 1/h^{2}$) of the stiffness
parameter (see Eq.(\ref{eta(h)t0})) further enhances the sheet resistance at
high fields by diminishing the localization length ($\xi \left( t\rightarrow
0\right) \propto 1/h\sqrt{\varepsilon _{h}}$).

\section{Quantum tunneling and pair breaking in the boson-insulating state}

It is evident that, in light of the limited number of unpaired electrons
available for the total conductivity, the ultimately divergent negative
conductance implied by Eq.(\ref{sig^fluctdt0}) is an unphysical result,
which clearly indicates the nature of the correction introduced in Ref.\cite%
{MZPRB2021}. In particular, the limitlessly rising Cooper-pairs density
within mesoscopic puddles, predicted by Eq.(\ref{sig^fluctdt0}) in the zero
temperature limit, can be stopped only by allowing the superfluous Cooper
pairs to tunnel out of the puddles while breaking into unpaired mobile
electron states. This should prevent the vanishing of the total conductivity
and the consequent divergence of the sheet resistance at high fields.

The formal incorporation of such a quantum correction into the thermal
conductance fluctuation, which was made in Ref. \cite{MZPRB2021}, amounts to
multiplying the AL term in Eq.(\ref{sig^fluctdlin}) by the factor $\left(
1+T_{Q}/T\right) $, where $T_{Q}$ stands for the tunneling attempt rate, and
dividing the DOS conductivity term by the same factor (see Appendix B for
the physical motivation). In parallel with these external corrections, the
electron pairing functions $\varepsilon _{H}$ and $\eta \left( H\right) $
appearing in Eq.(\ref{sig^fluctdlin}) were modified by inserting the
frequency-shift term $T_{Q}/2T$\ to the arguments of the digamma functions
and their derivatives in Eq.(\ref{eps_H}) and Eq.(\ref{eta}) respectively
(see Appendix B for more details). The external corrections are equivalent
to replacing the stiffness parameter appearing in the prefactors of the AL
and the DOS terms in Eq.(\ref{sig^fluctdlin}) with the hybrid expression: 
\begin{equation}
\eta \left( H\right) \rightarrow \left( 1+\frac{T_{Q}}{T}\right) \eta
_{U}\left( h\right)  \label{eta_correct}
\end{equation}%
where $\eta _{U}\left( h\right) $ is obtained from $\eta \left( h\right) $
in Eq.(\ref{eta(h)}) by inserting, under the Fermion Matsubara frequency
summation, the frequency-shift term $T_{Q}/2T$ : {\footnotesize 
\begin{align}
& \eta _{U}\left( h\right) =\sum\limits_{n=0}^{\infty }  \label{eta_U(h)} \\
& \frac{\left( n+1/2+T_{Q}/2T+2\beta +\overline{\delta }h^{2}\right) ^{2}-%
\overline{\mu }^{2}h^{2}}{\left[ \left( n+1/2+T_{Q}/2T+\overline{\delta }%
h^{2}\right) \left( n+1/2+T_{Q}/2T+2\beta +\overline{\delta }h^{2}\right) +%
\overline{\mu }^{2}h^{2}\right] ^{2}}  \notag
\end{align}%
}

In Eq.(\ref{eta_correct}), $\left( {\normalsize 1+T_{Q}/T}\right) $\
represents the effect of quantum tunneling of Cooper pairs, whereas the
frequency-shift term $T_{Q}/2T$ appearing in Eq.(\ref{eta_U(h)}) for $\eta
_{U}\left( h\right) $, represents pair-breaking effect associated with the
tunneling process.

The corresponding pair-breaking effect on the critical-shift parameter
results in transforming $\varepsilon _{H}$ according to: 
\begin{eqnarray}
\varepsilon _{H} &\rightarrow &\varepsilon _{h}^{U}\equiv \ln \left( \frac{T%
}{T_{c0}}\right) +a_{+}\psi \left( \frac{1}{2}+T_{Q}/2T+f_{-}\right)
\label{eps_Hcorr} \\
&&+a_{-}\psi \left( \frac{1}{2}+T_{Q}/2T+f_{+}\right) -\psi \left( 1/2\right)
\notag
\end{eqnarray}

In the absence of quantum tunneling $\varepsilon _{H}$ (Eq.(\ref{eps_H})) is
subjected to the usual magnetic field induced pair-breaking effect \cite%
{ShahLopatin07} through either the Zeeman spin-splitting energy ($\mu _{B}H$%
),\ or/and the diamagnetic energy ($D\left( deH\right) ^{2}/\hslash $)
terms. In the zero temperature limit, the effect is dramatically reflected
in the removal of the (Cooper) singularity of the logarithmic term in Eq.(%
\ref{eps_H}), due to exact cancellation by the asymptotic values of the
digamma functions for $\ f_{\pm }\gg 1$ (see Appendix C). In the presence of
quantum tunneling, the excitation frequency shift $\pi k_{B}$$T_{Q}/\hbar $\
introduced to define $\varepsilon _{h}^{U}$, Eq.(\ref{eps_Hcorr}), causes in
this limit an additional, field-independent pair-breaking effect through the
asymptotic behavior of the digamma functions for $T_{Q}/2T\gg 1$\ (see
Appendix C).

For systems with long range phase coherence described, e.g. in Ref.\cite%
{ShahLopatin07}, \cite{Lopatinetal05} the main impact of the pair-breaking
perturbations is near the critical point $\varepsilon _{H}=0$ for Cooper
pairs condensation (at $q=0)$ in momentum space. For the system of strong SC
fluctuations at very low temperatures, under consideration here, where
Cooper pairs tend to condense within mesoscopic puddles in real space, and
their excitation process associated with the frequency shift $\pi
k_{B}T_{Q}/\hbar $ \ greatly disperse in momentum space, dynamical quantum
tunneling, which is inherently connected to this excitation process,
strongly reinforces pair-breaking processes into unpaired electron states.

The sharp plunge of $\eta \left( h\right) $ just above $h=0$ for $%
T\rightarrow 0$, discussed below Eq.(\ref{eta(h)t0}) (see Fig.1), is a
reflection in the stiffness parameter of the field-induced pair-breaking
effect. As indicated above, the frequency shift that transforms $\eta \left(
h\right) $ to $\eta _{U}\left( h\right) $, and represents field-independent
pair breaking effect, is intimately connected to the quantum tunneling
process discussed above. This is clearly seen by considering the zero
temperature limit of $\eta _{U}\left( h\right) $ in Eq.(\ref{eta_U(h)}):

\begin{equation}
\left( \eta _{U}\left( h\right) \right) _{T\rightarrow 0}=\left( \frac{T}{%
T_{Q}}\right) _{T\rightarrow 0}\eta _{Q}\left( h\right)  \label{QLeta_U}
\end{equation}%
where: {\small 
\begin{eqnarray}
&&\eta _{Q}\left( h\right) \equiv  \notag \\
&& \int\limits_{0}^{\infty }d\nu \frac{\left( \nu +1/2+2\beta _{Q}+\delta
_{Q}h^{2}\right) ^{2}-\mu _{Q}^{2}h^{2}}{\left[ \left( \nu +1/2+\delta
_{Q}h^{2}\right) \left( \nu +1/2+2\beta _{Q}+\delta _{Q}h^{2}\right) +\mu
_{Q}^{2}h^{2}\right] ^{2}}  \notag  \label{eta_Q(h)} \\
&&=\frac{1}{h^{2}}\frac{\delta _{Q}h^{2}+2\beta _{Q}}{h^{2}\delta
_{Q}^{2}+2\beta _{Q}\delta _{Q}+\mu _{Q}^{2}}
\end{eqnarray}%
} and: $\beta _{Q}=\beta _{0}/t_{Q},\mu _{Q}=\mu _{0}/t_{Q},\delta
_{Q}=\delta _{0}/t_{Q},t_{Q}\equiv T_{Q}/T_{c}^{\ast }$.

The limiting function $\eta _{Q}\left( h\right) $ in Eq.(\ref{eta_Q(h)}) is
a continuous smooth function of the field $h$, including at $h=0$. \
Therefore, Eq.(\ref{QLeta_U}) implies that the discontinuous plunge of $\eta
\left( h\right) $ at $h=0$ in the zero temperature limit is removed by the
frequency shift term, as can be directly checked in Eq.(\ref{eta_U(h)}). The
overall magnitude of $\eta _{U}\left( h\right) $ diminishes to zero with $%
T/T_{Q}$ in this limit. However, by multiplying with the divergent quantum
tunneling factor $\left( 1+T_{Q}/T\right) $ the resulting hybrid product in
Eq.(\ref{eta_correct}), which represents the combined effect of quantum
tunneling and pair breaking, is a smooth finite function of the field $\eta
_{Q}\left( h\right) $ (see Fig.1).

\begin{figure}[tbh]
\includegraphics[width =.45\textwidth]{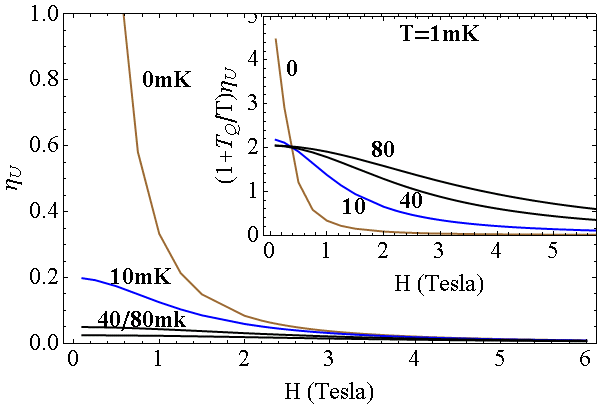}
\caption{Field-dependent stiffness parameter $\protect\eta _{U}\left(
H\right) $ calculated at $T=1$ mK for $T_{Q}=0,10,40,80$ mK. Inset: The
hybrid product $\left( 1+T_{Q}/T\right) \protect\eta _{U}\left( H\right) $
calculated for the same $T$ and $T_{Q}$ values as presented in the main
figure.}
\end{figure}

Our self-consistent field (SCF) approach, exploited in Ref.\cite{MZPRB2021}
for calculating the critical-shift parameter $\widetilde{{\normalsize %
\varepsilon }}_{H}$ in the presence of interaction between Gaussian
fluctuations, avoids the critical divergence of both the AL paraconductivity
and the DOS conductivity, and allows to extend Eq.(\ref{sig^fluctdlin}) for
the conductance fluctuations to regions well below the nominal critical SC
transition. It also offers an extended proper measure of the{\normalsize \ }%
pair-breaking effect. In contrast to $\varepsilon $$_{H}$, $\widetilde{%
{\normalsize \varepsilon }}_{H}$ is positive definite in the entire fields
range, including that below the critical field where $\varepsilon $$_{H}<0$
(see Fig.2). The uniform enhancement of $\widetilde{{\normalsize \varepsilon 
}}$$_{H}^{U}$ with respect to $\widetilde{{\normalsize \varepsilon }}_{H}$,
seen in Fig.2, resulting from the introduction of the frequency shift to the
SCF equation (see Ref.\cite{MZPRB2021}), is a genuine measure of the
pair-breaking effect associated with the frequency shift. Its monotonically
increasing field dependence seen in Fig.2 properly reflects the
field-induced pair-breaking effect in the entire fields range.

\begin{figure}[tbh]
\includegraphics[width =.45\textwidth]{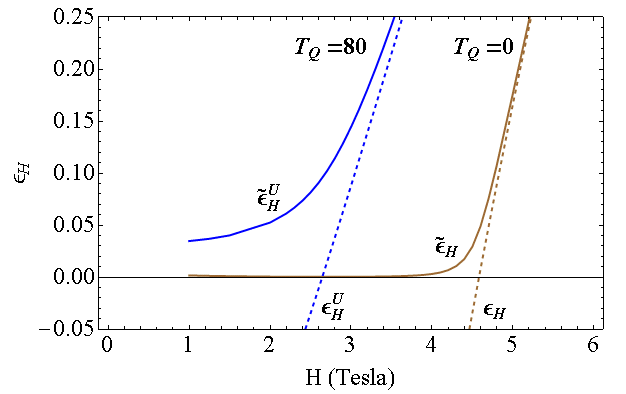}
\caption{Field dependence, at $T=30$ mK, of the "bare" critical-shift
parameter $\protect\varepsilon _{H}^{U}$ (dashed lines), and the
corresponding self-consistently "dressed" parameter $\widetilde{\protect%
\varepsilon }_{H}^{U}$ (solid lines), in the absence of quantum tunneling
(brown curves) and for $T_{Q}=80$ mK (blue curves). \ Note the downward
shift of the critical field and the uniform enhancement of the dressed
critical-shift parameter associated with the quantum tunneling effect.}
\end{figure}

\section{Sensitivity tests of the fitting process}

Practically speaking, the quantum tunneling introduced into the thermal
fluctuations theory--an essential requirement for avoiding the unphysical
divergence of the high-field DOS conductivity at zero temperature, and the
normal-electron conductivity term, which is closely related to the
pair-breaking processes bound to the tunneling events, are both
phenomenological constituents of our model, which are exclusively determined
by the experimental sheet-resistance data reported in Ref.\cite{Mograbi19}.

The other parameters in this model have microscopic origins and so can
either be evaluated from first principles or be extracted independently from
(other) experiments. Among the former group of microscopic parameters the
numerical prefactor of the total fluctuations conductance given by Eq.(\ref%
{sig^fluctdlin}) can be checked versus the relevant literature \cite{LV05}.
At zero field, where the stiffness parameter $\eta \left( h=0\right) =\pi
^{2}/2$, is independent of the spin-orbit energy parameter $\beta _{0}$, and
the corresponding conductance is:

{\normalsize 
\begin{equation}
\sigma ^{fluct}\left( H=0\right) \simeq \left( \frac{e^{2}}{\hbar d}\right) %
\left[ \left( \frac{1}{8}\right) \frac{1}{\varepsilon }-\left( \frac{7\zeta
\left( 3\right) }{\pi ^{4}}\right) \ln \left( \frac{\pi ^{2}x_{c}}{%
2\varepsilon }\right) \right]  \label{sig^fluctdh0}
\end{equation}%
}the prefactor is found here to be twice larger than that reported in Ref.%
\cite{LV05}. While we have not been able so far to successfully trace back
to the origin of this discrepancy such a variation in the amplitude of the
total (AL plus DOS) fluctuations conductivity is not expected to
significantly change the results of the fitting process presented in Ref. 
\cite{MZPRB2021}. As will be elaborated below, the results of the fitting
process can exactly be reproduced by slightly readjusting only the two
phenomenological parameters of the theory:--the tunneling attempt rate, $%
T_{Q}\left( T,H\right) $, and the normal-state conductivity, $\sigma
_{n}\left( H,T\right) .$

An important parameter in the fitting process is the magnetic field $H_{\max
}$ at which the sheet resistance has its maximum:-- the outstanding feature
characterizing the emergence of the insulating state at low temperatures.
This parameter is predominantly determined by the location of the minimum $%
H_{\min }$ of the fluctuations conductivity $\sigma ^{fluct}\left(
H,T\right) $, but is slightly shifted downward due to the field dependence
(increasing with increasing field) of the normal state conductivity 
{\normalsize $\sigma _{n}$}$\left( H,T\right) $. In the absence of quantum
tunneling, $\sigma ^{fluct}\left( H,T\right) $ at very low temperature
exhibits an asymmetrical sharp minimum arising from the opposing effects of
the sharply diminishing AL term with increasing field above the nominal
(mean-field) critical point and the less sharply decreasing DOS conductivity
term in Eq.(\ref{sig^fluctdlin}). The relevant field dependencies of these
terms above the nominal critical point are controlled by the field
dependencies of {\normalsize $\eta \left( h\right) $ }and $\varepsilon $$%
_{h} $, as shown in Figs.1 and 2, respectively. The dimensionless spin-orbit
energy parameter $\beta _{0}$ exclusively determines $H_{\min }$ in the
absence of quantum tunneling. The dependence of $H_{\max }$ on the gate
voltage shown in Fig.3 is therefore conveyed through the dependence of $%
\beta _{0}$ on the Fermi energy $E_{F}$.

Allowing for quantum tunneling of Cooper pairs, with attempt rate $T_{Q}$,
the sharp minimum of $\sigma ^{fluct}\left( H,T\right) $ is smeared, and due
to the asymmetry of the latter, $H_{\min }$ is shifted downward (see Fig.4).
The corresponding shift of $H_{\max }$, in conjunction with the downward
shift associated with the field dependence of {\normalsize $\sigma _{n}$}$%
\left( H,T\right) $, enable us fitting the data by exclusively varying the
phenomenological parameters $T_{Q}\left( H,T\right) $ and {\normalsize $%
\sigma _{n}$}$\left( H,T\right) $, without changing the other parameters.
Note that, in contrast to $H_{\min }$ which shifts downward with decreasing $%
\beta _{0}$ (or $E_{F}$), the depths of the minima in Fig.4 are seen to be
independent of $\beta _{0}$. $\ $Consequently, the high-field tails of $%
\sigma ^{fluct}\left( H,T\right) $ for the smaller value of $\beta _{0}$
shown in Fig.4, which always lay below zero, are seen to situate above the
corresponding tails for the larger value of $\beta _{0}$. To compensate for
these negative values, the field independent normal-state conductivity
parameter in our fitting process, $\sigma _{0}\left( T\right) $, should be
smaller for smaller values of $\beta _{0}$ (as indeed found, see Table I).
This feature reflects the underlying consistency of our fluctuations
approach (through the dependence of $\sigma ^{fluct}\left( H,T\right) $ upon
the normal-state carrier density) with the experimentally observed
high-field resistance (through its dependence on the gate voltage).

\begin{figure}[tbh]
\includegraphics[width =.45\textwidth]{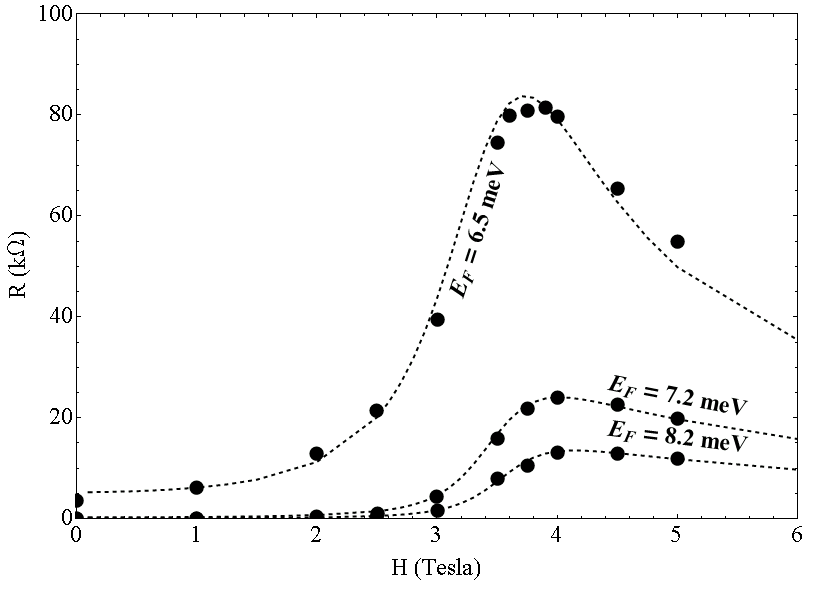}
\caption{Measured sheet resistance at $T=30$ mK, as a function of field for
three gate voltages (corresponding to $R_{N}=20.5,10.5,7.5$ k$\Omega $) as
reported in Ref. \protect\cite{Mograbi19} (full circles). The dashed lines
represent the results of calculations (for the respective Fermi energies: $%
E_{F}=6.5,7.2,8.2$ meV) similar to those performed in Ref. \protect\cite%
{MZPRB2021} , but with $1/2$ of the total amplitude of $\protect\sigma %
^{fluct}\left( H,T\right) $, and modified phenomenological adjustable
parameters $T_{Q}\left( H,T\right) $ and {\protect\normalsize $\protect%
\sigma _{n}$}$\left( H,T\right) $, as described in the text.}
\end{figure}

The results of the fittings of the sheet resistance data at $T=30$mK for the
various gate voltages, using the reduced total amplitude of $\sigma
_{U}^{fluct}\left( H,T\right) $, is shown in Fig.3. The quality of the
agreement with the experimental data is identical to that found in Ref.\cite%
{MZPRB2021} with the larger amplitude. The values of the phenomenological
fitting parameters obtained for the smaller amplitude are given in Table I
in Appendix D. \ All the other (i.e. microscopic) parameters have not been
changed. Variations of the tunneling attempt rate for the two amplitudes and
two gate voltages in the entire temperatures range are shown in Fig.5. The
collapse of all values of $T_{Q}\left( H_{\max },T\right) $ shown in Fig.5
in the zero temperature limit to: $T_{Q}\left( H_{\max },T\rightarrow
0\right) =60$ mK, reflects some sort of universality of the quantum
tunneling phenomenon which requires further investigation.

\bigskip

\section{Discussion}

The model system, introduced in Ref.\cite{Mograbi19} and further analyzed in
the present paper, has been motivated by the experimental observations of
pronounced MR peaks above a crossover field to superconductivity \cite%
{Mograbi19},\cite{MZPRB2021} in the high mobility electron systems formed in
the electron-doped SrTiO$_{3}$/LaAlO$_{3}$ (111) interface. Similar
electrostatically tuned SIT was reported for the LaAlO$_{3}$/SrTiO$_{3}$
(001) interface \cite{Caviglia08}, showing however \cite{Biscaras13},\cite%
{Mehta14} no clear indication of pronounced MR peaks similar to those
reported for the (111) interface. The theory predicts great sensitivity of
the fluctuation-induced MR peaks, observed at high fields, to variation of
the electronic interface density of states in the transition region of
strong spin-orbit induced band-mixing \cite{DiezPRL15}, \cite{KhannaPRL19}, 
\cite{JoshuaNcomm12} (see Fig.3). This feature was exploited in Ref.\cite%
{MZPRB2021}{\normalsize \ }for extracting the mobile electrons states
density from the experimental sheet-resistance data just by varying the gate
voltage. The sensitivity of the fitting process to uncertainty in the
overall amplitude of the conductance fluctuations has been tested in the
present paper, showing only minor changes restricted to the phenomenological
parameters, i.e. the normal state conductivity and the quantum-tunneling
attempt rate (see Appendix D and Fig.5).

For the various 2D electrons' systems generated by varying the gate voltage
applied to this interface, the diminishing sheet resistance measured at
decreasing temperature down to $30$ mK in the low magnetic fields region,
has not reached the ultimate zero-resistance characterizes a genuine SC
state. The self-consistent treatment of the interaction between fluctuations 
\cite{UllDor90} employed in our analysis accounts well for this feature and
for the consequent absence of a true critical point, allowing to extend the
theory to regions well below the nominal critical SC transition.

In our search for the deep origin of the high-field insulating states we
have discovered that, under increasing magnetic field, Cooper-pair
fluctuations in the zero temperature limit tend to localize within
mesoscopic puddles of decreasing spatial size, $\xi \left( t\rightarrow
0\right) =\left( \eta _{0}\left( h\right) /h^{2}\varepsilon _{h}\right)
^{1/2}\left( \hbar D/4\pi k_{B}T_{c}^{\ast }\right) ^{1/2}$ while developing
an infinitely large mass. The emerging picture of condensation of
Cooper-pairs in real space puddles is of course ideal, but basically
reflects real tendency toward a boson insulating state. It also calls for a
pair-breaking mechanism into unpaired electron states, stimulated by quantum
tunneling of Cooper pairs, which prevents the unphysical divergence of the
Cooper-pairs density.

\begin{figure}[tbh]
\includegraphics[width =.45\textwidth]{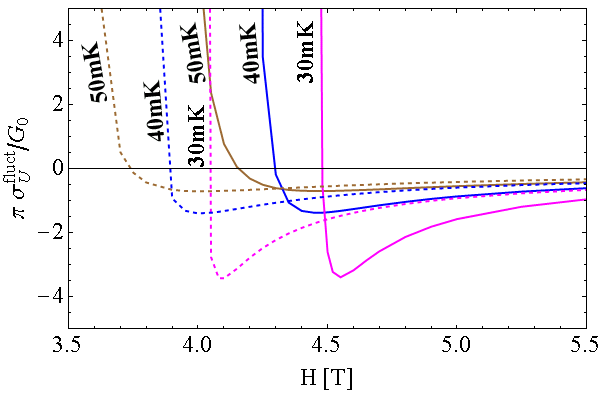}
\caption{Normalized fluctuations conductivity, $\protect\pi \protect\sigma %
_{U}^{fluct}\left( H,T\right) /G_{0}$, as a function of field at $T=1$ mK,
around its minimum, for $\protect\beta _{0}=14$ (solid lines) and for $%
\protect\beta _{0}=11$ (dashed lines), for three values of the quantum
tunneling attempt rate; $T_{Q}=30$ mK (magenta), $T_{Q}=40$ mK (blue) and $%
T_{Q}=50$ mK (brown). Note the downward shifts of $H_{\min }$ with the
decreasing values of $\protect\beta _{0}$ and/or the increasing values of $%
T_{Q}$.}
\end{figure}

Realization of this scenario in 2D electron systems with strong spin-orbit
scatterings under a parallel magnetic field at low temperatures shows that
at sufficiently high fields the DOS conductivity prevails over the
paraconductivity, resulting in strongly enhanced MR in systems with
sufficiently small carriers density. Dynamical quantum tunneling of Cooper
pairs, breaking into mobile normal-electrons states, contain the resistance
onset at high magnetic field. In this system of heavy, charged bosons in
equilibrium with unpaired mobile electrons, the dilute system of mobile
electrons are responsible for most of the residual conductance.

\begin{figure}[tbh]
\includegraphics[width =.45\textwidth]{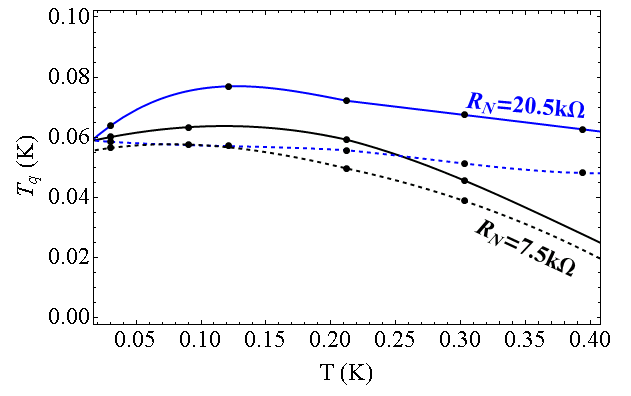}
\caption{Best fitting values of the quantum tunneling attempt rate $%
T_{Q}\left( H_{\max },T\right) $ at the maximum point $H_{\max }$ of the
sheet resistance as a function of Temperatures, for $\protect\beta _{0}=11$ (%
$R_{N}=20.5$ k$\Omega $, blue lines), and for $\protect\beta _{0}=14$ ($%
R_{N}=7.5$ k$\Omega $, black lines). \ Dashed lines represent results for
the fitting expression employed in Ref.\protect\cite{MZPRB2021} , whereas
the solid lines represent the results for the same expression, in which the
total amplitude of $\protect\sigma _{U}^{fluct}\left( H,T\right) $ was
multiplied by 1/2. }
\end{figure}

An important feature of the localization process predicted in this approach
is its dynamical nature, namely that it{\normalsize \ }occurs in response to
the driving electric force \cite{MZPRB2021}, and not spontaneously in a
thermodynamical process toward equilibrium state. This feature seems to
distinguish it from the various approaches to the phenomenon of SIT
discussed in the literature \cite{Dubi07}, \cite{Bouadim2011}, \cite%
{GhosalPRL98}, \cite{Vinokur2008}, in which disorder-induced spatial
inhomogeneity in the form of SC islands is involved in generating the
insulating state. However, in a similar manner the formation of fluctuation
puddles in our approach is controlled by disorder, which strongly affect the
Cooper-pairs amplitude correlation function in real space. This can be seen
by comparing the pair correlation function derived in the dirty limit \cite%
{MZPRB2021},\cite{CaroliMaki67I} to that obtained in the pure limit \cite%
{CaroliMaki67II}.

Another important parameter in our approach of relevance to the insulating
behavior that seems to have a parallel in the literature \cite{Bouadim2011},
is the self-consistent critical shift parameter $\widetilde{\varepsilon }%
_{H} $, which also plays the role of an energy gap in the Cooper-pair
fluctuations spectrum \cite{MZPRB2021}. Thus, it is interesting to note that
the two-particle gap, which characterizes the insulating state in Ref.\cite%
{Bouadim2011}, vanishes at the SIT. Analogously, in our approach the
(two-particle) Cooper-pair fluctuation gap $\widetilde{\varepsilon }_{H}$
gradually diminishes to very small (nonvanishing)\ values upon decreasing
field below the sheet-resistance peak (see Fig.2 and Fig.3), in accord with
the lack of a critical point.

\bigskip

\section{Acknowledgments}

We would like to thank Eran Maniv, Itai Silber and Yoram Dagan for helpful
discussions.

\appendix

\section{The DOS conductivity from microscopic theory}

In this appendix we evaluate the DOS conductivity in a 2D system in the zero
field limit, following the fully microscopic (diagrammatic) approach
presented in Ref.\cite{LV05} for a layered superconductor.

Starting with diagram No.5, and using the notation employed in Ref.\cite%
{LV05} (according to which $\hbar =k_{B}=1$ and the distance between layers
is $s$) the corresponding response function is given by:

\begin{equation*}
Q_{xx}^{\left( 5\right) }\left( \omega \right) =i\omega \kappa _{1}\left(
T\tau \right) \frac{\pi \eta _{\left( 2\right) }e^{2}}{4s}A_{xx}\frac{1}{%
\left( 2\pi \right) ^{2}}\int \frac{d^{2}q}{\varepsilon +\eta _{\left(
2\right) }q^{2}}
\end{equation*}%
where $\varepsilon \equiv \ln \left( T/T_{c0}\right) $, $A_{xx}=2\left%
\langle v^{2}\right\rangle _{FS}/v_{F}^{2}=2\left\langle \cos ^{2}\theta
\right\rangle =1$, and $\eta _{\left( 2\right) }=\pi D/8T$, so that by
performing the integration over $x\equiv \eta _{\left( 2\right) }q^{2}$,
i.e.: $Q_{xx}^{\left( 5\right) }\left( \omega \right) =i\omega \kappa
_{1}\int_{0}^{x_{c}}dx\left( \varepsilon +x\right) ^{-1}\left(
e^{2}/16s\right) $, one finds:

{\normalsize 
\begin{equation*}
Q_{xx}^{\left( 5\right) }\left( \omega \right) =i\omega \kappa _{1}\frac{%
e^{2}}{16s}\ln \left( \frac{x_{c}}{\varepsilon }\right)
\end{equation*}%
}

The corresponding conductivity:

\begin{equation}
\sigma _{xx}^{\left( 5\right) }=-\frac{Q_{xx}^{\left( 5\right) }\left(
\omega \right) }{i\omega }=-\kappa _{1}\frac{e^{2}}{16s}\ln \left( \frac{%
x_{c}}{\varepsilon }\right)  \label{sig^(5)}
\end{equation}%
which together with the topologically equivalent diagram 6 gives:

{\normalsize 
\begin{equation*}
\sigma _{xx}^{\left( 5+6\right) }=2\sigma _{xx}^{\left( 5\right) }=-\kappa
_{1}\frac{e^{2}}{8s}\ln \left( \frac{x_{c}}{\varepsilon }\right)
\end{equation*}%
}

For the two other diagrams 7, and 8, the result is:

{\normalsize 
\begin{equation*}
\sigma _{xx}^{\left( 7+8\right) }=2\sigma _{xx}^{\left( 7\right) }=-\kappa
_{2}\frac{e^{2}}{8s}\ln \left( \frac{x_{c}}{\varepsilon }\right)
\end{equation*}%
}

Taking into account all the four diagrams contributing to the DOS
conductivity we have for the 2D limit:

\begin{equation*}
\sigma _{xx}^{\left( 5+6+7+8\right) }=-\frac{e^{2}}{8s}\kappa \left( T\tau
\right) \ln \left( \frac{x_{c}}{\varepsilon }\right)
\end{equation*}%
where: 
\begin{equation*}
\kappa \left( T\tau \right) \equiv \kappa _{1}+\kappa _{2}=\frac{-\psi
^{\prime }\left( \frac{1}{2}+\frac{1}{4\pi T\tau }\right) +\frac{1}{2\pi
T\tau }\psi ^{\prime \prime }\left( \frac{1}{2}\right) }{\pi ^{2}\left[ \psi
\left( \frac{1}{2}+\frac{1}{4\pi T\tau }\right) -\psi \left( \frac{1}{2}%
\right) -\frac{1}{4\pi T\tau }\psi ^{\prime }\left( \frac{1}{2}\right) %
\right] }
\end{equation*}%
\newline

Estimating $\kappa \left( T\tau \right) $ in the dirty limit: $T\tau \ll 1$
by exploiting the asymptotic expansion of the digamma function, $\psi \left(
z\right) \rightarrow \ln z,\psi ^{\prime }\left( z\right) \rightarrow 1/z$,
we find: $\ \kappa \left( T\tau \right) _{T\tau \ll 1}\rightarrow -2\psi
^{\prime \prime }\left( \frac{1}{2}\right) /\pi ^{2}\psi ^{\prime }\left( 
\frac{1}{2}\right) ,\psi ^{\prime }\left( \frac{1}{2}\right) =\pi
^{2}/2,\psi ^{\prime \prime }\left( z\right) =-14\zeta \left( 3\right) $, so
that: $\kappa \left( T\tau \right) _{T\tau \ll 1}\rightarrow 8\times 7\zeta
\left( 3\right) /\pi ^{4}$, and:

{\normalsize 
\begin{eqnarray}
&&\left( \sigma _{xx}^{\left( 5+6+7+8\right) }\right) _{T\tau \ll
1}\rightarrow -\kappa \left( T\tau \right) _{T\tau \ll 1}\frac{e^{2}}{8s}\ln
\left( \frac{x_{c}}{\varepsilon }\right)  \notag \\
&=&-\left( \frac{7\zeta \left( 3\right) }{\pi ^{4}}\right) \left( \frac{e^{2}%
}{s}\right) \ln \left( \frac{x_{c}}{\varepsilon }\right)  \label{sigDOS_LV}
\end{eqnarray}%
}

It should be stressed at this point that this expression, which was derived
here directly from the 2D limit of the response function $Q_{xx}\left(
\omega \right) $, as presented in Ref.\cite{LV05} for a layered (quasi 2D)
system, is by a factor of 2 larger than the 2D limit of the final expression
for the total DOS conductivity reported in Ref.\cite{LV05}.

\section{ The quantum fluctuations correction to conductivity}

In this appendix we outline the physical reasoning behind our
phenomenological quantum fluctuations correction to the two ingredients of
the conductance fluctuations. Starting with the DOS conductivity we consider
the Cooper-pair density, $n_{s}$, given in Eq.(\ref{n_s}), with $%
\left\langle \left\vert \psi \left( q\right) \right\vert ^{2}\right\rangle $
in Eq.(\ref{MomDistr}). Approximating $7\zeta \left( 3\right) \simeq 8.4$ we
rewrite:

\begin{equation}
\left\langle \left\vert \psi \left( q\right) \right\vert ^{2}\right\rangle
\simeq \left( \frac{2.1E_{F}}{\pi ^{2}k_{B}T}\right) \frac{1}{\Phi \left(
x;\varepsilon _{H}\right) }=4.2\frac{\left( N_{2D}\lambda _{T}^{2}\right) }{%
\Phi \left( x;\varepsilon _{H}\right) }  \label{MomDist}
\end{equation}%
where $N_{2D}=k_{F}^{2}/2\pi $ is the density of the 2D electron gas and $%
\lambda _{T}=\sqrt{\hbar ^{2}/2\pi m^{\ast }k_{B}T}$ is the thermal
wavelength.

The momentum distribution function $\left\langle \left\vert \psi \left(
q\right) \right\vert ^{2}\right\rangle $ measures the number of bosons per
wave vector $\mathbf{q}$ in the Cooper-pairs liquid, engaged in equilibrium
with a 2D gas of unpaired mobile electrons with a nominal density $N_{2D}$.
The prefactor $N_{2D}\lambda _{T}^{2}=\left( 1/2\pi ^{2}\right) \left(
E_{F}\tau _{T}/\hbar \right) $, that is the number of electrons in an area
of size equal to the thermal wavelength, is proportional to the
characteristic thermal activation time $\tau _{T}=\hbar /k_{B}T$.

The quantum corrections, introduced in Ref.\cite{MZPRB2021}, amount to
modifying Expression \ref{MomDist} in two steps; in the first, replacing the
temperature $T$, appearing in the denominator of the prefactor, with $%
T+T_{Q} $, and in the second step inserting the frequency-shift term $%
T_{Q}/2T$\ to the arguments of the digamma functions in Eq.(\ref{Phi(x)})
consistently with the replacement of $\varepsilon _{H}$ with $\varepsilon
_{H}^{U}$. The total modification takes the form:

\begin{eqnarray*}
\left\langle \left\vert \psi \left( q\right) \right\vert ^{2}\right\rangle
&\rightarrow& \left\langle \left\vert \psi _{U}\left( q\right) \right\vert
^{2}\right\rangle =N_{2D}\lambda _{U}^{2}\frac{4.2}{\Phi _{U}\left(
x;\varepsilon _{H}^{U}\right) }  \notag \\
&=&\frac{2.1}{\pi ^{2}}\frac{\left( E_{F}\tau _{U}/\hbar \right) }{\Phi
_{U}\left( x;\varepsilon _{H}^{U}\right) }
\end{eqnarray*}%
where $1/\tau _{U}=1/\tau _{T}+1/\tau _{Q}$, and $\tau _{Q}=\hbar
/k_{B}T_{Q} $, is the characteristic time for Cooper-pair tunneling. The
prefactor $N_{2D}\lambda _{U}^{2}$, is the number of electrons in an
effective area $\lambda _{U}^{2}=\hbar ^{2}/2\pi m^{\ast }k_{B}\left(
T+T_{Q}\right) $ that is proportional to the characteristic time, $\tau _{U}$%
, for both thermal activation and quantum tunneling of Cooper pairs. Thus,
increasing the temperature and/or shortening the time $\tau _{Q}$ for
quantum tunneling (which also enhance pair breaking by increasing $\Phi
_{U}\left( x;\varepsilon _{H}^{U}\right) $), result in larger rate of
thermal and/or quantum leakage from puddles of Cooper pairs. The resulting
reduction in the number of Cooper-pairs, which occurs versus a corresponding
increase in the number of unpaired mobile electrons, would suppress the DOS
contribution to the resistance.

The corresponding unified (quantum thermal (QT)) density (per unit area) of
the Cooper-pairs liquid is now evaluated: $n_{s}^{U}=\frac{1}{d}\frac{1}{%
\left( 2\pi \right) ^{2}}\int \left\langle \left\vert \psi _{U}\left(
q\right) \right\vert ^{2}\right\rangle d^{2}q=\frac{1}{d}\frac{1}{\left(
2\pi \right) ^{2}}\int_{0}^{q_{c}^{2}}\pi d\left( q^{2}\right) \left( \frac{%
2.1E_{F}}{\pi ^{2}k_{B}\left( T+T_{Q}\right) }\right) \frac{1}{\Phi
_{U}\left( x;\varepsilon _{H}^{U}\right) }$, so that the unified DOS
conductivity, $\sigma _{DOS}^{U}=-\left( 2n_{s}^{U}e^{2}/m^{\ast }\right)
\tau _{SO}$, is given by:

\begin{equation}
\sigma _{DOS}^{U}d\simeq -4.2\left( \frac{G_{0}}{\pi }\right)
\int_{0}^{t^{-1}x_{0}}\frac{dx}{\left( 1+T_{Q}/T\right) \Phi _{U}\left(
x;\varepsilon _{H}^{U}\right) }  \label{sig^U_DOSd}
\end{equation}

For the AL thermal fluctuations conductivity we start with the retarded
current-current correlator $Q_{AL}^{R}\left( \omega \right) $, Eq.(\ref%
{Q_AL^R}), which was obtained from the Matsubara correlator $Q_{AL}\left(
i\Omega _{\nu }\right) $ following the analytic continuation $i\Omega _{\nu
}\rightarrow \omega $. \ The corresponding electrical response function is
seen to be proportional to the thermal energy $k_{B}T=\hbar /\tau _{T}$. \
The effects of quantum tunneling and pair breaking are introduced by adding
to the thermal attempt rate $1/\tau _{T}\propto k_{B}T$ the quantum
tunneling attempt rate $1/\tau _{Q}\propto k_{B}T_{Q}$ , and by
appropriately inserting the frequency-shift term $T_{Q}/2T$ \ into the
function $\Phi \left( x+\left\vert n+y\right\vert ;\varepsilon _{H}\right) $%
, as explained in the main text, i.e.:

\begin{widetext}

\begin{equation*}
Q_{AL}^{U,R}\left( \omega \right) =k_{B}\left( T+T_{Q}\right) \left( \frac{2e%
}{\hbar }\right) ^{2}\left( \frac{1}{2\pi d}\right)
\int\limits_{0}^{x_{c}}xdx\sum\limits_{n=0,\pm 1,\pm 2,....}\frac{\Phi
_{U}^{\prime }\left( x+\left\vert n+y\right\vert ;\varepsilon
_{H}^{U}\right) }{\Phi _{U}\left( x+\left\vert n+y\right\vert ;\varepsilon
_{H}^{U}\right) }\frac{\Phi _{U}^{\prime }\left( x+\left\vert n\right\vert
;\varepsilon _{H}^{U}\right) }{\Phi _{U}\left( x+\left\vert n\right\vert
;\varepsilon _{H}^{U}\right) }
\end{equation*}%
where $2i\pi yk_{B}T/\hslash =\omega $.

The corresponding conductivity is: $\sigma _{AL}^{U}=\lim_{\omega
\rightarrow 0}\left( i/\omega \right) \left[ Q_{AL}^{U,R}\left( \omega
\right) -Q_{AL}^{U,R}\left( 0\right) \right] =$

$k_{B}\left( T+T_{Q}\right) \lim_{y\rightarrow 0}\left( -\frac{\hbar }{2\pi
k_{B}Ty}\right) \left( \frac{2e}{\hbar }\right) ^{2}\left( \frac{1}{2\pi d}%
\right) \int\limits_{0}^{\infty }xdx\sum\limits_{n=0,\pm 1,\pm 2,....}\left[ 
\frac{\Phi _{U}^{\prime }\left( x+\left\vert n+y\right\vert ;\varepsilon
_{H}^{U}\right) }{\Phi _{U}\left( x+\left\vert n+y\right\vert ;\varepsilon
_{H}^{U}\right) }\frac{\Phi _{U}^{\prime }\left( x+\left\vert n\right\vert
;\varepsilon _{H}^{U}\right) }{\Phi _{U}\left( x+\left\vert n\right\vert
;\varepsilon _{H}^{U}\right) }-\frac{\Phi _{U}^{\prime }\left( x+\left\vert
n\right\vert ;\varepsilon _{H}^{U}\right) }{\Phi _{U}\left( x+\left\vert
n\right\vert ;\varepsilon _{H}^{U}\right) }\frac{\Phi _{U}^{\prime }\left(
x+\left\vert n\right\vert ;\varepsilon _{H}^{U}\right) }{\Phi _{U}\left(
x+\left\vert n\right\vert ;\varepsilon _{H}^{U}\right) }\right] $

$=-\left( \frac{T+T_{Q}}{T}\right) \left( \frac{e}{\pi }\right) ^{2}\left( 
\frac{1}{d\hbar }\right) \int\limits_{0}^{\infty }xdx\sum\limits_{n=0,\pm
1,\pm 2,....}\frac{\Phi _{U}^{\prime }\left( x+\left\vert n\right\vert
;\varepsilon _{H}^{U}\right) }{\Phi _{U}\left( x+\left\vert n\right\vert
;\varepsilon _{H}^{U}\right) }\lim_{y\rightarrow 0}\frac{1}{y}\left[ \frac{%
\Phi _{U}^{\prime }\left( x+\left\vert n+y\right\vert ;\varepsilon
_{H}^{U}\right) }{\Phi _{U}\left( x+\left\vert n+y\right\vert ;\varepsilon
_{H}^{U}\right) }-\frac{\Phi _{U}^{\prime }\left( x+\left\vert n\right\vert
;\varepsilon _{H}^{U}\right) }{\Phi _{U}\left( x+\left\vert n\right\vert
;\varepsilon _{H}^{U}\right) }\right] $, \ which can be reduced to (compare
Eq.\ref{sig_ALd}):

\begin{equation}
\sigma _{AL}^{U}d=\frac{1}{4}\left( \frac{G_{0}}{\pi }\right) \left( 1+\frac{%
T_{Q}}{T}\right) \int\limits_{0}^{t^{-1}x_{0}}\left( \frac{\Phi _{U}^{\prime
}\left( x;\varepsilon _{H}^{U}\right) }{\Phi _{U}\left( x;\varepsilon
_{H}^{U}\right) }\right) ^{2}dx  \label{sig^U_ALd}
\end{equation}

\begin{table*}[tbp]
\resizebox{\textwidth}{!}{	\renewcommand{\arraystretch}{1.3} 	\begin{tabular}[t]{@{}ccccccccccccccccc}
		\multicolumn{5}{c}{$R_N =7.5 [k\Omega$]} & \phantom{abcd} & 
		\multicolumn{5}{c}{$R_N =10.5 [k\Omega$]} & \phantom{abcd} & 
		\multicolumn{5}{c}{$R_N =20.5 [k\Omega$]} \\ 
		\cmidrule{1-5} \cmidrule{7-11} \cmidrule{13-17} $T [mK]$ & $T_q [mK]$ & $H_q
		[T]$ & $H_n [T]$ & $\sigma_0[k\Omega^{-1}]$ &  & $T [mK]$ & $T_q [mK]$ & $		H_q [T]$ & $H_n [T]$ & $\sigma_0[k\Omega^{-1}]$ &  & $T[mK]$ & $T_q [mK]$ & $		H_q [T]$ & $H_n [T]$ & $\sigma_0[k\Omega^{-1}]$ \\ 
		\midrule 30 & 83 & 8 & 6.9 & .065 &  & 30 & 80 & 7.3 & 7 & .041 &  & 30 & 97
		& 6.5 & 4.25 & .014 \\ 
		90 & 77 & 10 & 7 & .065 &  & 130 & 75 & 10 & 7.1 & .041 &  & 121 & 90 & 10 & 
		4.35 & .014 \\ 
		212 & 62 & 20 & 12 & .09 &  & 230 & 62 & 15 & 8 & .051 &  & 212 & 82 & 10 & 
		4.5 & .015 \\ 
		303 & 47 & 25 & 14 & .096 &  & 330 & 55 & 18 & 10 & .065 &  & 303 & 72 & 12
		& 6.5 & .026 \\ 
		485 & 10 & 30 & 20 & .106 &  & 430 & 35 & 25 & 12 & .069 &  & 394 & 67 & 15
		& 8 & .029 \\ 
		\bottomrule &  &  &  &  &  &  &  &  &  &  &  &  &  &  &  & 
	\end{tabular}    }
\caption{Values of the temperature-dependent parameters extracted in the
fitting process of the measured sheet resistance for the 1/2-reduced
amplitude of $\protect\sigma ^{fluct}\left( H,T\right) $, which determine
the temperature and field dependencies of the phenomenological parameters $%
\protect\sigma _{n}\left( H,T\right) $ and $T_{Q}\left( T,H\right) $ (see
Appendix D for details). \ As elaborated in Ref. \protect\cite{MZPRB2021},
the three gate voltages employed correspond to $R_{N}=20.5,10.5,7.5$ k$%
\Omega $ .}
\label{table1}
\end{table*}

\end{widetext}

\section{The Quantum limit of the Critical-shift parameter}

In this appendix we study the pair-breaking effect due to magnetic field and
to quantum tunneling of Cooper pairs in the zero temperature limit. \
Consider the unified (quantum-thermal) expression, Eq.(\ref{eps_Hcorr}), for
the critical shift parameter $\varepsilon _{h}^{U}$ in the zero-temperature
(quantum) limit. \bigskip\ Using the asymptotic expansion of $\psi \left( 
\frac{1}{2}+T_{Q}/2T+f_{\pm }\right) $ for $T_{Q}/T,$ $f_{\pm }\gg 1$, i.e. $%
\psi \left( \frac{1}{2}+T_{Q}/2T+f_{\pm }\right) \rightarrow \ln \left(
T_{Q}/2T+f_{\pm }\right) =\ln \left[ \left( T_{Q}+T_{\pm }\right) /2T\right] 
$, we have: \ 
\begin{eqnarray}
&&\varepsilon _{h}^{U}\rightarrow \varepsilon _{h}^{Q}=\ln \left(
T/T_{c0}\right) -\ln T +  \label{eps_h^UT0} \\
&&a_{+}\ln \left( T_{Q}+T_{-}\right) +a_{-}\ln \left( T_{Q}+T_{+}\right)
-\ln 2-\psi \left( 1/2\right)  \notag
\end{eqnarray}%
where:

\begin{equation}
T_{\pm }\equiv \frac{D\left( de\right) ^{2}H^{2}}{\pi k_{B}\hslash }+\frac{%
\varepsilon _{SO}}{2\pi k_{B}}\pm \sqrt{\left( \frac{\varepsilon _{SO}}{2\pi
k_{B}}\right) ^{2}-\left( \frac{\mu _{B}H}{\pi k_{B}}\right) ^{2}}
\label{T_PM}
\end{equation}

In the above expression for $\varepsilon _{h}^{Q}$ (Eq.\ref{eps_h^UT0}), the
Cooper singular term, $\ln \left( T/T_{c0}\right) $, is exactly cancelled by
the logarithmic term arising from the asymptotic expansion of the digamma
functions, so that the remaining regular terms are rearranged to yield the
following temperature independent expression for ${\normalsize \varepsilon
_{h}^{Q}}$: 
\begin{equation}
{\normalsize \varepsilon _{h}^{Q}\rightarrow a_{+}\ln \left( \frac{%
T_{Q}+T_{-}}{T_{c0}}\right) +a_{-}\ln \left( \frac{T_{Q}+T_{+}}{T_{c0}}%
\right) +\ln 2+\gamma }  \label{epas_h^Q}
\end{equation}%
where $\gamma \approx 0.5772$... is the Euler--Mascheroni constant, and:

\begin{equation*}
a_{\pm }=\frac{1}{2}\left( 1\pm \frac{1}{\sqrt{1-\left( \mu _{0}/\beta
_{0}\right) ^{2}h^{2}}}\right)
\end{equation*}%
with $\mu _{0}\equiv \mu _{B}H_{c\parallel 0}^{\ast }/2\pi k_{B}T_{c}^{\ast
},\beta _{0}\equiv \varepsilon _{SO}/4\pi k_{B}T_{c}^{\ast }$.

\section{ The phenomenological fitting parameters}

As in our fitting process, described in Ref.\cite{MZPRB2021}, the
normal-state conductivity contribution $\sigma _{n}\left( H,T\right) $ has a
quadratic field-dependent form: $\sigma _{n}\left( H,T\right) =\sigma
_{0}\left( T\right) \left[ 1+\left( H/H_{n}\left( T\right) \right) \right]
^{2}$, with two adjustable, temperature-dependent parameters{\large \ }$%
\sigma _{0}\left( T\right) ,H_{n}\left( T\right) ${\large . }The
corresponding expression for the MR, defined as usual by: $MR\left(
H,T\right) \equiv \left[ \rho _{n}\left( H,T\right) -\rho _{n}\left(
0,T\right) \right] /\rho _{n}\left( 0,T\right) $, where $\rho _{n}\left(
H,T\right) =1/\sigma _{n}\left( H,T\right) $, is given by: 
\begin{equation}
MR\left( H,T\right) =-\frac{\left( H/H_{n}\left( T\right) \right) ^{2}}{%
1+\left( H/H_{n}\left( T\right) \right) ^{2}}  \label{MR(H,T)}
\end{equation}%
yielding negative MR in qualitative agreement with that observed in Refs.%
\cite{RoutPRB17} and \cite{DiezPRL15}{\large \ }at temperatures well above $%
T_{c}$.

Similarly, for the temperature and field dependence of the phenomenological
quantum tunneling "temperature" parameter $T_{Q}\left( T,H\right) $ we use
here the form employed in Ref.\cite{MZPRB2021}:

\begin{equation}
T_{Q}\left( T,H\right) =T_{Q}\left( T\right) \left[ 1-\left( \frac{H}{%
H_{Q}\left( T\right) }\right) ^{2}\right]  \label{T_Q(H,T)}
\end{equation}%
with the two adjustable parameters, $T_{Q}\left( T\right) $ and $H_{Q}\left(
T\right) $.

The best fitting values for $T_{Q}\left( T\right) ,H_{Q}\left( T\right)
,H_{n}\left( T\right) $, and $\sigma _{0}\left( T\right) $, obtained in our
fitting process for the 1/2-reduced amplitude of $\sigma ^{fluct}\left(
H,T\right) $ are listed in Table I.


\begin{thebibliography}{99}
\bibitem{MZPRB2021} T. Maniv and V. Zhuravlev, "Superconducting fluctuations
and giant negative magnetoresistance in a gate-voltage tuned two-dimensional
electron system with strong spin-orbit impurity scattering", Phys. Rev. B
104, 054503 (2021).

\bibitem{Ohtomo04} A. Ohtomo, and H. Y. Hwang, "A high-mobility electron gas
at the LaAlO$_{3}$/SrTiO$_{3}$ heterointerface", Nature 427, 423 (2004).

\bibitem{Mograbi19} M. Mograbi, E. Maniv, P. K. Rout, D. Graf, J. -H Park
and Y. Dagan, "Vortex excitations in the Insulating

State of an Oxide Interface", Phys. Rev. B 99, 094507 (2019).

\bibitem{UllDor90} S. Ullah and A.T. Dorsey, "Critical Fluctuations in
High-Temperature Superconductors and the Ettingshausen Effect", Phys. Rev.
Lett. 65, 2066 (1990). Properties of (111) LaAlO$_{3}$/SrTiO$_{3}$", Phys.
Re. Lett. 123, 036805 (2019).

\bibitem{UllDor91} S. Ullah and A.T. Dorsey, "Effect of fluctuations on the
transport properties of type-II superconductors in a magnetic field", Phys.
Rev. B 44, 262 (1991).

\bibitem{AL68} L. G. Aslamazov and A.I. Larkin, Phys. Lett. A 26 p. 238
(1968).

\bibitem{LV05} A. Larkin and A. Varlamov, "Theory of fluctuations in
superconductors", Oxford University Press 2005.

\bibitem{footnote1} In Ref.\cite{MZPRB2021} we have made two technical
errors, which nearly cancelled each other, while approximately arriving to
the exact\ expression \ref{sigDOS_LV} derived in Appendix A.

\bibitem{footnote2} Estimation of the argument of the logarithmic factor in
Eq.(\ref{sig^fluctdt0}) just above the "nominal" critical field $%
H_{c\parallel 0}^{\ast }=4.5$T ($\varepsilon _{h\gtrsim 1}=0.05$) in the $%
t\rightarrow 0$ limit, based on typical values of our fitting parameters
yields $\left( \eta _{0}\left( h\right) x_{0}/h^{2}\varepsilon _{h}\right)
_{h\gtrsim 1}\approx 1.3$. \ Estimations of the field-dependent prefactors
of the AL and the DOS conductivities under the same conditions yield,
respectively: $\left( \eta _{0}\left( h\right) /4h^{2}\right) _{h=1}\simeq
\left( 3.5\zeta \left( 3\right) h^{2}/\eta _{0}\left( h\right) \right)
_{h=1}\simeq 1$.

\bibitem{ShahLopatin07} N. Shah and A. V. Lopatin, "Microscopic analysis of
the superconducting quantum critical point: Finite-temperature crossovers in
transport near a pair-breaking quantum phase transition", Phys. Rev. B 76,
094511 (2007).

\bibitem{Lopatinetal05} A. V. Lopatin, N. Shah, and V. M. Vinokur,
"Fluctuation Conductivity of Thin Films and Nanowires Near a
Parallel-Field-Tuned Superconducting Quantum Phase Transition", Phys. Rev.
Lett. 94, 037003 (2005).

\bibitem{Caviglia08} A. D. Caviglia, S. Gariglio, N. Reyren, D. Jaccard, T.
Schneider, M. Gabay, S. Thiel, G. Hammerl, J. Mannhart and J.-M. Triscone,
"Electric field control of the LaAlO3/SrTiO3 interface ground state", Nature
(London) 456, 624 (2008).

\bibitem{Biscaras13} J. Biscaras, N. Bergeal, S. Hurand, C. Feuillet-Palma,
A. Rastogi, R. C. Budhani, M. Grilli, S. Caprara and J. Lesueur, "Multiple
quantum criticality in a two-dimensional superconductor", Nat. Mater. 12,
542 (2013).

\bibitem{Mehta14} M. M. Mehta, D. A. Dikin, C. W. Bark, S. Ryu, C. M.
Folkman, C. B. Eom, and V. Chandrasekhar, "Magnetic field tuned
superconductor-to-insulator transition at the LaAlO$_{3}$/SrTiO$_{3}$
interface", Phys. Rev. B 90, 100506 (2014).\ 

\bibitem{DiezPRL15} M. Diez, A. M. R. V. L. Monteiro, G. Mattoni, E.
Cobanera, T. Hyart, E. Mulazimoglu, N. Bovenzi, C.W. J. Beenakker, and A. D.
Caviglia, "Giant Negative Magnetoresistance Driven by Spin-Orbit Coupling at
the LaAlO$_{3}$/SrTiO$_{3}$ Interface", Phys. Re. Lett. 115, 016803 (2015).

\bibitem{KhannaPRL19} Udit Khanna, P. K. Rout, Michael Mograbi, Gal Tuvia,
Inge Leermakers, Uli Zeitler, Yoram Dagan, and Moshe Goldstein, "Symmetry
and Correlation Effects on Band Structure Explain the Anomalous Transport
Properties of (111) LaAlO$_{3}$/SrTiO$_{3}$", Phys. Re. Lett. 123, 036805
(2019).

\bibitem{JoshuaNcomm12} Arjun Joshua, S. Pecker, J. Ruhman, E. Altman and S.
Ilani, "A universal critical density underlying the physics of electrons at
the LaAlO$_{3}$/SrTiO$_{3}$ interface", Nat. Commun. 3, 1129 (2012).

\bibitem{Dubi07} Y. Dubi, Y. Meir and Y. Avishai, "Nature of the
superconductor--insulator transition in disordered superconductors", Nature
449, 876 (2007).

\bibitem{Bouadim2011} K. Bouadim, Y. L. Loh, M. Randeria and N. Trivedi,
"Single- and two-particle energy gaps across the disorder-driven
superconductor--insulator transition", Nature Phys. 7, 884\ (2011).

\bibitem{GhosalPRL98} A. Ghosal, M. Randeria, and N. Trivedi, "Role of
Spatial Amplitude Fluctuations in Highly Disordered s-Wave Superconductors",
Phys. Re. Lett. 81, 3940 (1998).

\bibitem{Vinokur2008} V. Vinokur, T. I. Baturina, M. V. Fistul, A. Yu.
Mironov, M. R. Baklanov and C. Strunk, "Superinsulator and quantum
synchronization", Nature 452, 613 (2008). \ 

\bibitem{CaroliMaki67I} C. Caroli and K. Maki, "Fluctuations of the Qrder
Parameter in Type-II Superconductors. I. Dirty Limit", Phys. Rev. 159, 306
(1967).

\bibitem{CaroliMaki67II} C. Caroli and K. Maki, "Fluctuations of the Qrder
Parameter in Type-II Superconductors. II. Pure Limit", Phys. Rev. 159, 316
(1967).

\bibitem{RoutPRB17} P. K. Rout, I. Agireen, E. Maniv, M. Goldstein, and Y.
Dagan, "Six-fold crystalline anisotropic magnetoresistance in the (111) LaAlO%
$_{3}$/SrTiO$_{3}$ oxide interface", Phys. Rev. B \textbf{95}, 241107(R)
(2017).
\end{thebibliography}
\end{document}